\documentclass[twocolumn,showpacs,amsmath,amssymb]{revtex4}

\usepackage{graphicx}

\begin{document}
\title{
Binegativity and geometry of entangled states in two qubits}
\author{Satoshi Ishizaka}
%\email{isizaka@frl.cl.nec.co.jp}
\affiliation{
PRESTO, JST, 4-1-8 Honcho Kawaguchi, Saitama, Japan. \\
Fundamental Research Laboratories, NEC Corporation, \\
34 Miyukigaoka, Tsukuba, Ibaraki, 305-8501 Japan}
\date{\today}
%%%%%%%%%%%%%%%%%%%%%%%%%%%%%%%%%%%%%%%%%%%%%%%%%%%%%%%%%%%%%%%%%%%%%%%%%%%%%%%
\begin{abstract}
We prove that the binegativity
is always positive for any two-qubit state.
As a result, as suggested by the previous works, 
the asymptotic relative entropy of
entanglement in two qubits does not exceed the Rains bound,
and the PPT-entanglement cost
for any two-qubit state is determined to be the logarithmic negativity
of the state.
Further, the proof reveals some geometrical characteristics of the entangled
states, and shows that the partial transposition can give another separable
approximation of the entangled state in two qubits.
\end{abstract}
%%%%%%%%%%%%%%%%%%%%%%%%%%%%%%%%%%%%%%%%%%%%%%%%%%%%%%%%%%%%%%%%%%%%%%%%%%%%%%%
\maketitle
%%%%%%%%%%%%%%%%%%%%%%%%%%%%%%%%%%%%%%%%%%%%%%%%%%%%%%%%%%%%%%%%%%%%%%%%%%%%%%%
Quantum entanglement plays an essential role in many quantum information tasks,
and qualitative and quantitative understanding of the entanglement is
one of the central topics in quantum information theory.
An important mathematical operation in the theory of entanglement
is the partial transposition $\sigma^{T_B}$ \cite{Peres96a},
where only the basis on one party, say Bob, is transposed.
The states satisfying $\sigma^{T_B}\!\ge\!0$ are called positive partial
transposed (PPT) states, and all separable states are PPT states.
Further, it has been shown that all PPT states in $2\!\otimes\!2$ (two qubits) and
$2\!\otimes\!3$ are separable states \cite{Horodecki96a}.
\par
%%%%%%%%%%%%%%%%%%%%%%%%%%%%%%%%%%%%%%%%%%%%%%%%%%%%%%%%%%%%%%%%%%%%%%%%%%%%%%%
Recently, Audenaert, Moor, Vollbrecht
and Werner introduced an interesting and important mathematical
operation, the binegativity $|\sigma^{T_B}|^{T_B}$ \cite{Audenaert02a}.
They showed that, if $|\sigma^{T_B}|^{T_B}\!\ge\!0$ holds,
the asymptotic relative entropy of entanglement with respect to
PPT states does not exceed the so-called Rains bound.
Further they showed that $|\sigma^{T_B}|^{T_B}\!\ge\!0$ holds
for many classes of states and conjectured that it holds for
any two-qubit state.
Subsequently, Audenaert, Plenio, and Eisert showed that,
if $|\sigma^{T_B}|^{T_B}\!\ge\!0$ holds,
the PPT-entanglement cost for the exact preparation
is given by the logarithmic negativity \cite{Audenaert03a}. 
By this, they provided an operational meaning to the logarithmic
negativity.
\par
%%%%%%%%%%%%%%%%%%%%%%%%%%%%%%%%%%%%%%%%%%%%%%%%%%%%%%%%%%%%%%%%%%%%%%%%%%%%%%%
In this paper, we prove that $|\sigma^{T_B}|^{T_B}\!\ge\!0$ indeed
holds for any two-qubit state.
The proof is geometrical in some sense, and reveals some geometrical
characteristics of the entangled states.
Further, it is found that the partial transposition can give
another separable approximation of the entangled state in two qubits.
\par
%%%%%%%%%%%%%%%%%%%%%%%%%%%%%%%%%%%%%%%%%%%%%%%%%%%%%%%%%%%%%%%%%%%%%%%%%%%%%%%
Before starting the proof of $|\sigma^{T_B}|^{T_B}\!\ge\!0$,
we briefly review several concepts necessary to the proof.
The first is the entanglement witness
\cite{Horodecki96a,Terhal00a,Lewenstein00a},
which is the Hermitian operator $W$ such that $\hbox{Tr}W\varrho\!\ge\!0$
for all separable states $\varrho$, and $\hbox{Tr}W\sigma\!<\!0$ for some
entangled states $\sigma$.
This is expressed such that $W$ {\it detects} the entanglement of $\sigma$.
For $\sigma^{T_B}\!\not\ge\!0$,
$(|\psi\rangle\langle\psi|)^{T_B}$ is the entanglement
witness where $|\psi\rangle$ is the eigenstate of $\sigma^{T_B}$
for a negative eigenvalue \cite{Guhne02a}.
Since the entanglement witness cannot be a positive operator
\cite{Lewenstein00a},
$|\psi\rangle$ must always be entangled.
\par
%%%%%%%%%%%%%%%%%%%%%%%%%%%%%%%%%%%%%%%%%%%%%%%%%%%%%%%%%%%%%%%%%%%%%%%%%%%%%%%
The concept of the entanglement witness is related to the
existence of the hyper-plane separating the closed convex set of
separable states and some entangled states.
$W$ itself plays the role of the normal vector of the
hyper-plane.
The concept can be applicable to another closed convex set of
positive operators:
if $\hbox{Tr}\sigma|\phi\rangle\langle\phi|\!<\!0$,
it is certain that $\sigma$ is not positive.
In this paper, by the analogy to the entanglement witness, we call
$|\phi\rangle\langle\phi|$ witness of the nonpositivity,
and we say that
$|\phi\rangle\langle\phi|$ {\it detects} the nonpositivity of $\sigma$.
\par
%%%%%%%%%%%%%%%%%%%%%%%%%%%%%%%%%%%%%%%%%%%%%%%%%%%%%%%%%%%%%%%%%%%%%%%%%%%%%%%
The second concept relates to the state representation based on the local
filtering.
According to Ref. \cite{Verstraete01a},
all states in two qubits can be transformed 
by local filtering of full rank into either
the Bell diagonal states or the states of
%%%%%%%%%%%%%%%%%%%%%%%%%%%%%%%%%%%%%%%%%%%%%%%%%%%%%%%%%%%%%%%%%%%%%%%%%%%%%%%
\begin{equation}
\sigma_c=\frac{1}{2}
\left(
\begin{array}{cccc}
a+c & 0 &  0  & d \\
 0  & 0 &  0  & 0 \\
 0  & 0 & b-c & 0 \\
 d  & 0 &  0  & a-b \\
\end{array}
\right)
\label{eq: sigma_c}
\end{equation}
%%%%%%%%%%%%%%%%%%%%%%%%%%%%%%%%%%%%%%%%%%%%%%%%%%%%%%%%%%%%%%%%%%%%%%%%%%%%%%%
with $a$, $b$, $c$ and $d$ being real.
Therefore, all states in two qubits can be represented by either
%%%%%%%%%%%%%%%%%%%%%%%%%%%%%%%%%%%%%%%%%%%%%%%%%%%%%%%%%%%%%%%%%%%%%%%%%%%%%%%
\begin{equation}
\sigma=\frac{1}{N}\sum_{i=0}^3 p_i
(A \otimes B)|e_i\rangle\langle e_i|(A^\dagger \otimes B^\dagger)
\label{eq: normal form I} 
\end{equation}
%%%%%%%%%%%%%%%%%%%%%%%%%%%%%%%%%%%%%%%%%%%%%%%%%%%%%%%%%%%%%%%%%%%%%%%%%%%%%%%
or
%%%%%%%%%%%%%%%%%%%%%%%%%%%%%%%%%%%%%%%%%%%%%%%%%%%%%%%%%%%%%%%%%%%%%%%%%%%%%%%
\begin{equation}
\sigma=\frac{1}{N}
(A \otimes B)\sigma_c(A^\dagger \otimes B^\dagger),
\label{eq: normal form II} 
\end{equation}
%%%%%%%%%%%%%%%%%%%%%%%%%%%%%%%%%%%%%%%%%%%%%%%%%%%%%%%%%%%%%%%%%%%%%%%%%%%%%%%
where $N$ is the normalization,
$\sum_i p_i\!=\!1$, and $|e_i\rangle$ is the set of orthogonal Bell basis.
Without loss of generality,
we can fix 
$|e_i\rangle\!=\!
\{|\psi^-\rangle,|\psi^+\rangle,|\phi^-\rangle,|\phi^+\rangle\}$
\cite{Note1,Horodecki96b},
where $|\psi^\pm\rangle\!=\!(|01\rangle\!\pm\!|10\rangle)/\sqrt{2}$
and $|\phi^\pm\rangle\!=\!(|00\rangle\!\pm\!|11\rangle)/\sqrt{2}$,
and we can assume that $p_0$ is largest and $p_3$ is smallest among
$p_i$ \cite{Note1,Horodecki96b,Bennett96a}.
Further, since $A$ and $B$ are full rank, we can put
$\det A\!=\!\det B\!=\!1$ without loss of generality.
This leads to a convenient relation of
$A^\dagger \tilde A\!=\!B^\dagger \tilde B\!=\!I$
\cite{Verstraete01a,Cen02a},
where the tilde operation is defined as
$\tilde A\!\equiv\!\sigma_2A^*\sigma_2$ for local operators
and $|\tilde \psi\rangle\!\equiv\!(\sigma_2\!\otimes\!\sigma_2)|\psi^*\rangle$
for states \cite{Wootters98a}.
In this paper, these forms of the state representation
[Eqs.\ (\ref{eq: normal form I}) or (\ref{eq: normal form II})]
are called normal form.
\par
%%%%%%%%%%%%%%%%%%%%%%%%%%%%%%%%%%%%%%%%%%%%%%%%%%%%%%%%%%%%%%%%%%%%%%%%%%%%%%%
Then, let us start the proof of $|\sigma^{T_B}|^{T_B}\!\ge\!0$.
It is trivial when $\sigma$ is a PPT state, 
since $|\sigma^{T_B}|^{T_B}\!=\!\sigma\!\ge\!0$.
Therefore, we restrict ourselves to the case that $\sigma$ is entangled.
The partial transposition of $\sigma$ can be written as
%%%%%%%%%%%%%%%%%%%%%%%%%%%%%%%%%%%%%%%%%%%%%%%%%%%%%%%%%%%%%%%%%%%%%%%%%%%%%%%
\begin{equation}
\sigma^{T_B}=P-\lambda|\psi\rangle\langle\psi|,
\label{eq: P}
\end{equation}
%%%%%%%%%%%%%%%%%%%%%%%%%%%%%%%%%%%%%%%%%%%%%%%%%%%%%%%%%%%%%%%%%%%%%%%%%%%%%%%
where $|\psi\rangle$ is the (normalized) eigenstate for the
negative eigenvalue of $-\lambda$
(there is only one negative eigenvalue for entangled
states in two qubits \cite{Verstraete01d}).
The remainder $P$ is the (unnormalized) positive part ($P\!\ge\!0$),
which is orthogonal to $|\psi\rangle$, and hence $P|\psi\rangle\!=\!0$.
Then, $\sigma$ and $|\sigma^{T_B}|^{T_B}$ are
%%%%%%%%%%%%%%%%%%%%%%%%%%%%%%%%%%%%%%%%%%%%%%%%%%%%%%%%%%%%%%%%%%%%%%%%%%%%%%%
\begin{eqnarray}
\sigma&=&P^{T_B}-\lambda(|\psi\rangle\langle\psi|)^{T_B}, \cr
|\sigma^{T_B}|^{T_B}&=&P^{T_B}+\lambda(|\psi\rangle\langle\psi|)^{T_B}.
\end{eqnarray}
%%%%%%%%%%%%%%%%%%%%%%%%%%%%%%%%%%%%%%%%%%%%%%%%%%%%%%%%%%%%%%%%%%%%%%%%%%%%%%%
\par
%%%%%%%%%%%%%%%%%%%%%%%%%%%%%%%%%%%%%%%%%%%%%%%%%%%%%%%%%%%%%%%%%%%%%%%%%%%%%%%
%%%%%%%%%%%%%%%%%%%%%%%%%%%%%%%%%%%%%%%%%%%%%%%%%%%%%%%%%%%%%%%%%%%%%%%%%%%%%%%
\begin{figure}
\centerline{\scalebox{0.45}[0.45]{\includegraphics{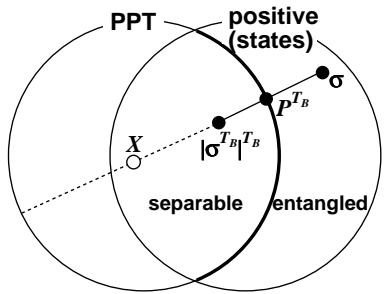}}}
\caption{
Schematic picture of set of positive operators (quantum states) and
set of PPT operators.
$\sigma$ and $|\sigma^{T_B}|^{T_B}$ are symmetrically located with
respect to $P^{T_B}$.
}
\label{fig: Positive and PPT ball}
\end{figure}
%%%%%%%%%%%%%%%%%%%%%%%%%%%%%%%%%%%%%%%%%%%%%%%%%%%%%%%%%%%%%%%%%%%%%%%%%%%%%%%
Here, it is worth discussing the geometrical meaning of the problem.
Let us consider the space of all Hermitian operators.
The set of quantum states which are positive operators is a subset of the
whole (we do not care about the normalization explicitly).
Further, let us consider PPT operators, which are those
such that its partial transposed operators are positive
(a PPT operator can be either positive or nonpositive).
These two sets are schematically shown in 
Fig.\ \ref{fig: Positive and PPT ball}
(this figure is essentially the same as Fig.\ 1 in Ref.\ \cite{Verstraete01c}).
Hereafter, we call these sets
{\it positive ball} and {\it PPT ball}, although the actual
shape is not a spherical ball \cite{Verstraete01c}.
The intersection of the two balls corresponds to the set of PPT states
(separable states).
Therefore, entangled state $\sigma$ is located in the positive ball
outside the intersection.
Since $|\sigma^{T_B}|^{T_B}$ is a PPT operator
[$(|\sigma^{T_B}|^{T_B})^{T_B}\!=\!|\sigma^{T_B}|\!\ge\!0$],
it is contained in the PPT ball.
Then, the geometrical meaning to prove $|\sigma^{T_B}|^{T_B}\!\ge\!0$ is
to prove that $|\sigma^{T_B}|^{T_B}$ is always located in the intersection.
\par
%%%%%%%%%%%%%%%%%%%%%%%%%%%%%%%%%%%%%%%%%%%%%%%%%%%%%%%%%%%%%%%%%%%%%%%%%%%%%%%
Further, let us pay attention to the geometrical location of $P^{T_B}$:
it is located on the middle of the line connecting
$\sigma$ and $|\sigma^{T_B}|^{T_B}$ since
$P^{T_B}\!=\!\sigma/2\!+\!|\sigma^{T_B}|^{T_B}/2$.
In addition, $P^{T_B}$ must be located on the edge of the PPT ball,
since $P^{T_B}$ itself is a PPT operator and its partial transposition 
is rank deficient ($P|\psi\rangle\!=\!0$).
However, so that $|\sigma^{T_B}|^{T_B}$ is located in the intersection,
it is geometrically obvious that $P^{T_B}$ must be located on the edge
of the intersection (thick solid curve in 
Fig.\ \ref{fig: Positive and PPT ball}).
This corresponds to $P^{T_B}\!>\!0$, which is indeed
necessary for $|\sigma^{T_B}|^{T_B}\!\ge\!0$ because 
$|\sigma^{T_B}|^{T_B}\!=\!2P^{T_B}\!-\!\sigma$.
\par
%%%%%%%%%%%%%%%%%%%%%%%%%%%%%%%%%%%%%%%%%%%%%%%%%%%%%%%%%%%%%%%%%%%%%%%%%%%%%%%
Then, the outline of the proof is as follows:
We first prove a lemma which simplifies the proof of
$|\sigma^{T_B}|^{T_B}\!\ge\!0$.
Second, we prove that $P^{T_B}\!>\!0$ (positive definite)
whenever a given state $\sigma$ is entangled.
Finally, we search for an operator $X$ in the intersection
such that $|\sigma^{T_B}|^{T_B}$ is located on the
line connecting $P^{T_B}$ and $X$ (see Fig.\ \ref{fig: Positive and PPT ball}).
As a result, it is found that $|\sigma^{T_B}|^{T_B}$ can be always written as
a convex sum of two positive operators ($P^{T_B}$ and $X$), 
which can complete the proof.
\par
%%%%%%%%%%%%%%%%%%%%%%%%%%%%%%%%%%%%%%%%%%%%%%%%%%%%%%%%%%%%%%%%%%%%%%%%%%%%%%%
The key point of the proof for $|\sigma^{T_B}|^{T_B}\!\ge\!0$ is to represent
$P$ (not $\sigma$) in the normal form mentioned before.
A lemma we first prove is concerned with the state representation of 
$P$.
\par
%%%%%%%%%%%%%%%%%%%%%%%%%%%%%%%%%%%%%%%%%%%%%%%%%%%%%%%%%%%%%%%%%%%%%%%%%%%%%%%
{\bf Lemma 1.}
{\it Let $\sigma$ be any entangled state in two qubits and
write down as $\sigma^{T_B}\!=\!P\!-\!\lambda|\psi\rangle\langle\psi|$
where $P\!\ge\!0$ and  $P|\psi\rangle\!=\!0$.
If $P$ is rank 3, $P$ is always represented in the normal form of
Eq.\ (\ref{eq: normal form I}).
If the rank of $P$ is less than 3, 
there always exist $\sigma'$ in the vicinity of $\sigma$ such that
$\sigma'^{T_B}\!=\!P'\!-\!\lambda|\psi\rangle\langle\psi|$
where $P'|\psi\rangle\!=\!0$, $P'$ is rank 3,
and $P'$ is represented in the normal form of
Eq.\ (\ref{eq: normal form I}).}
\par
%%%%%%%%%%%%%%%%%%%%%%%%%%%%%%%%%%%%%%%%%%%%%%%%%%%%%%%%%%%%%%%%%%%%%%%%%%%%%%%
{\it Proof:}
In the case that $P$ is rank 3,
let us assume that $P$ is represented in the normal
form of Eq.\ (\ref{eq: normal form II}) as
$P\!=\!\frac{1}{N}(A \otimes B)\sigma_c(A^\dagger \otimes B^\dagger)$
where $\sigma_c$ is given by Eq.\ (\ref{eq: sigma_c}).
Since $P$ is assumed to be rank 3 and $A\!\otimes\!B$ is full rank,
$\sigma_c$ must be rank 3.
By using $A^\dagger\tilde A\!=\!B^\dagger\tilde B\!=\!I$,
it can be easily checked that only the state of
$|\psi\rangle\!=\!\textstyle\frac{1}{\sqrt{M}}(\tilde A\!\otimes\!\tilde B)
|01\rangle$
%%%%%%%%%%%%%%%%%%%%%%%%%%%%%%%%%%%%%%%%%%%%%%%%%%%%%%%%%%%%%%%%%%%%%%%%%%%%%%%
satisfies $P|\psi\rangle\!=\!0$ where $M$ is the normalization.
However, this state is a product state which contradicts
that $|\psi\rangle$ must be an entangled state in order that
$(|\psi\rangle\langle\psi|)^{T_B}$ is an entanglement witness
detecting the entanglement of $\sigma$
(entanglement witness cannot be a positive operator as mentioned before).
Therefore, $P$ of rank 3 must be represented in the normal
form of Eq.\ (\ref{eq: normal form I}) as
%%%%%%%%%%%%%%%%%%%%%%%%%%%%%%%%%%%%%%%%%%%%%%%%%%%%%%%%%%%%%%%%%%%%%%%%%%%%%%%
\begin{equation}
P=\frac{1}{N}\sum_{i=0}^3
p_i (A \otimes B)|e_i\rangle\langle e_i|(A^\dagger \otimes B^\dagger),
\label{eq: Normal Form}
\end{equation}
%%%%%%%%%%%%%%%%%%%%%%%%%%%%%%%%%%%%%%%%%%%%%%%%%%%%%%%%%%%%%%%%%%%%%%%%%%%%%%%
where $p_3\!=\!0$ in order that $P$ is rank 3 since $p_3$ is smallest
among $p_i$.
Further, by fixing the Bell basis as
$|e_i\rangle\!=\!\{|\psi^-\rangle,|\psi^+\rangle,|\phi^-\rangle,
|\phi^+\rangle\}$,
it is found that only the state of
%%%%%%%%%%%%%%%%%%%%%%%%%%%%%%%%%%%%%%%%%%%%%%%%%%%%%%%%%%%%%%%%%%%%%%%%%%%%%%%
\begin{equation}
|\psi\rangle=\textstyle\frac{1}{\sqrt{M}}(\tilde A\otimes \tilde B)
|\phi^+\rangle
\label{eq: Kernel}
\end{equation}
%%%%%%%%%%%%%%%%%%%%%%%%%%%%%%%%%%%%%%%%%%%%%%%%%%%%%%%%%%%%%%%%%%%%%%%%%%%%%%%
satisfies $P|\psi\rangle\!=\!0$.
\par
%%%%%%%%%%%%%%%%%%%%%%%%%%%%%%%%%%%%%%%%%%%%%%%%%%%%%%%%%%%%%%%%%%%%%%%%%%%%%%%
In the case that the rank of $P$ is less than 3,
there always exist $P'$ of rank 3 in the vicinity of $P$
such that $P'|\psi\rangle\!=\!0$
(for example, if $P$ is rank 2,
using $|\psi^\perp\rangle$
orthogonal to $|\psi\rangle$ and satisfying $P|\psi^{\perp}\rangle\!=\!0$,
let
$P'\!=\!P+\epsilon|\psi^{\perp}\rangle\langle\psi^{\perp}|$
with $\epsilon$ being an infinitesimally small positive value).
This $P'$ must be represented in the normal form of
Eq.\ (\ref{eq: normal form I}) for the same reason discussed above
($|\psi\rangle$ satisfying $P'|\psi\rangle\!=\!0$ is entangled).
Since $P'$ is in the vicinity of $P$,
$\sigma'\!=\!P'^{T_B}\!-\!\lambda(|\psi\rangle\langle\psi|)^{T_B}$
is also in the vicinity of $\sigma$.
\hfill \mbox{\vrule width .6em height .6em}
\par
%%%%%%%%%%%%%%%%%%%%%%%%%%%%%%%%%%%%%%%%%%%%%%%%%%%%%%%%%%%%%%%%%%%%%%%%%%%%%%%
It should be noted that the rank of $P$ will be shown to be 3, and the
possibility of the second
case in Lemma 1 will be denied (see Corollary 1 below).
\par
%%%%%%%%%%%%%%%%%%%%%%%%%%%%%%%%%%%%%%%%%%%%%%%%%%%%%%%%%%%%%%%%%%%%%%%%%%%%%%%
\begin{figure}
\centerline{\scalebox{0.44}[0.44]{\includegraphics{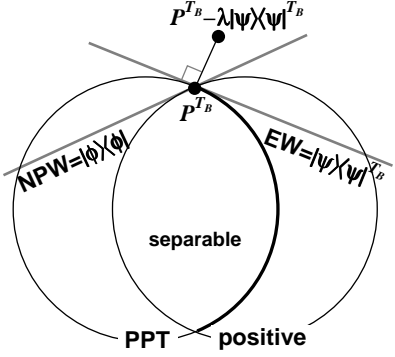}}}
\caption{
$P^{T_B}$ is located at crossing point of two edges.
Two hyper-planes of entanglement witness (EW)
and nonpositivity witness (NPW) are indicated by gray lines.
}
\label{fig: Crossing Point}
\end{figure}
%%%%%%%%%%%%%%%%%%%%%%%%%%%%%%%%%%%%%%%%%%%%%%%%%%%%%%%%%%%%%%%%%%%%%%%%%%%%%%%
The next task for the proof of $|\sigma^{T_B}|^{T_B}\!\ge\!0$ is to
prove $P^{T_B}\!>\!0$ whenever $\sigma$ is entangled.
To this end, it suffices to show that,
if we assume $P^{T_B}\!\not >\!0$,
$\sigma\!=\!P^{T_B}\!-\!\lambda(|\psi\rangle\langle\psi|)^{T_B}$
cannot be any entangled state for $\lambda\!>\!0$ and for $|\psi\rangle$
satisfying $P|\psi\rangle\!=\!0$.
According to Lemma 1, so that $\sigma$ is an entangled state,
$P$ (or $P'$ in the close vicinity of $P$) must be written
as Eq.\ (\ref{eq: Normal Form}) at least.
For those $P$ (or $P'$) of rank 3, Eq.\ (\ref{eq: Kernel}) is only the state
satisfying $P|\psi\rangle\!=\!0$ (or $P'|\psi\rangle\!=\!0$).
Therefore, in the following, we shall show that,
if $P^{T_B}\!\not>\!0$,
$\sigma\!=\!P^{T_B}\!-\!\lambda(|\psi\rangle\langle\psi|)^{T_B}$
cannot be positive (and hence cannot be an entangled state)
for every $P$ of Eq.\ (\ref{eq: Normal Form}) and for $|\psi\rangle$ of
Eq.\ (\ref{eq: Kernel}).
By this, when the rank of $P$ is less than 3,
since $\sigma'\!=\!P'^{T_B}\!-\!\lambda(|\psi\rangle\langle\psi|)^{T_B}$
cannot be positive as well, $\sigma$ in the close vicinity of
$\sigma'$ cannot be positive.
\par
%%%%%%%%%%%%%%%%%%%%%%%%%%%%%%%%%%%%%%%%%%%%%%%%%%%%%%%%%%%%%%%%%%%%%%%%%%%%%%%
The partial transposition of Eq.\ (\ref{eq: Normal Form})
is calculated as
%%%%%%%%%%%%%%%%%%%%%%%%%%%%%%%%%%%%%%%%%%%%%%%%%%%%%%%%%%%%%%%%%%%%%%%%%%%%%%%
\begin{eqnarray}
P^{T_B}&\!\!\!=\!\!\!&
\frac{1}{N}\sum_{i=0}^3
p_i(A\otimes B^*)(|e_i\rangle\langle e_i|)^{T_B}(A^\dagger \otimes B^T) \cr
&\!\!\!=\!\!\!&
\frac{1}{2N}\sum_{i=0}^3
(1-2p_{3-i})(A\otimes B^*)|e_i\rangle\langle e_i|(A^\dagger \otimes B^T),
\label{eq: PT of Normal Form}
\end{eqnarray}
%%%%%%%%%%%%%%%%%%%%%%%%%%%%%%%%%%%%%%%%%%%%%%%%%%%%%%%%%%%%%%%%%%%%%%%%%%%%%%%
where we fixed $|e_i\rangle$ as
$\{|\psi^-\rangle,|\psi^+\rangle,|\phi^-\rangle,|\phi^+\rangle\}$ and used
%%%%%%%%%%%%%%%%%%%%%%%%%%%%%%%%%%%%%%%%%%%%%%%%%%%%%%%%%%%%%%%%%%%%%%%%%%%%%%%
\begin{eqnarray}
(|\psi^\pm\rangle\langle\psi^\pm|)^{T_B}
=\textstyle\frac{1}{2}(
   |\psi^-\rangle\langle\psi^-|
&\!\!\!+\!\!\!&|\psi^+\rangle\langle\psi^+| \cr
\mp|\phi^-\rangle\langle\phi^-|
&\!\!\!\pm\!\!\!&|\phi^+\rangle\langle\phi^+|), \cr
(|\phi^\pm\rangle\langle\phi^\pm|)^{T_B}
=\textstyle\frac{1}{2}\big(
\mp|\psi^-\rangle\langle\psi^-|
&\!\!\!\pm\!\!\!&|\psi^+\rangle\langle\psi^+| \cr
+|\phi^-\rangle\langle\phi^-|
&\!\!\!+\!\!\!&|\phi^+\rangle\langle\phi^+|).
\end{eqnarray}
%%%%%%%%%%%%%%%%%%%%%%%%%%%%%%%%%%%%%%%%%%%%%%%%%%%%%%%%%%%%%%%%%%%%%%%%%%%%%%%
Further, $P^{T_B}\!\not>\!0$ corresponds
to $p_0\!\ge\!1/2$ in Eq.\ (\ref{eq: Normal Form}),
since $1\!-\!2p_0$ is smallest among $1\!-\!2p_i$
($P^{T_B}$ is positive semidefinite for $p_0\!=\!1/2$, and has a
negative eigenvalue for $p_0\!>\!1/2$).
By introducing the state of
%%%%%%%%%%%%%%%%%%%%%%%%%%%%%%%%%%%%%%%%%%%%%%%%%%%%%%%%%%%%%%%%%%%%%%%%%%%%%%%
\begin{equation}
|\phi\rangle=\textstyle\frac{1}{\sqrt{L}}(\tilde A\otimes \tilde B^*)|\phi^+\rangle,
\label{eq: Witness}
\end{equation}
%%%%%%%%%%%%%%%%%%%%%%%%%%%%%%%%%%%%%%%%%%%%%%%%%%%%%%%%%%%%%%%%%%%%%%%%%%%%%%%
where $L$ is the normalization,
it is found that
%%%%%%%%%%%%%%%%%%%%%%%%%%%%%%%%%%%%%%%%%%%%%%%%%%%%%%%%%%%%%%%%%%%%%%%%%%%%%%%
\begin{eqnarray}
\langle\phi|\sigma|\phi\rangle&=&\langle\phi|P^{T_B}|\phi\rangle
-\lambda\langle\phi|(|\psi\rangle\langle\psi|)^{T_B}|\phi\rangle \cr
&=&\frac{1-2p_0}{2NL}-\lambda 
\hbox{Tr}|\phi\rangle\langle\phi|(|\psi\rangle\langle\psi|)^{T_B} \cr
&=&\frac{1-2p_0}{2NL}-\frac{\lambda}{2ML}\hbox{Tr}
(C\otimes I) V (C^\dagger\otimes I) P^+ \cr
&=&\frac{1-2p_0}{2NL}-\frac{\lambda}{4ML}\hbox{Tr}CC^*,
\label{eq: Expectation}
\end{eqnarray}
%%%%%%%%%%%%%%%%%%%%%%%%%%%%%%%%%%%%%%%%%%%%%%%%%%%%%%%%%%%%%%%%%%%%%%%%%%%%%%%
where $V\!=\!2(|\phi^+\rangle\langle\phi^+|)^{T_B}$ is the flip operator,
and $C\!\equiv H_1 H_2$ is the product of two positive definite
operators
$ H_1\!\equiv\!\tilde A^\dagger \tilde A$ and
$ H_2\!\equiv\!
(\tilde B^\dagger \tilde B)^*\!=\!\tilde B^T \tilde B^*$.
In the third equality, we used 
$(A\!\otimes\!B)V(A^\dagger\!\otimes\!B^\dagger)\!=\!
(AB^\dagger\!\otimes\!I)V(BA^\dagger\!\otimes\!I)$.
Using $\det H_1\!=\!\det \tilde H_2\!=\! \det C\!=\!1$, it can be shown that
$\hbox{Tr}CC^* \!\ge\! 2$ \cite{Note2}.
As a result, 
it is found that $\langle\phi|\sigma|\phi\rangle\!<\!0$ for $\lambda\!>\!0$,
and $\sigma$ cannot be positive where it is assumed that
$P^{T_B}\!\not>\!0$.
In this way, $|\phi\rangle\langle\phi|$ works as a witness operator
detecting the nonpositivity of $\sigma$.
Then, the following theorem was proven.
\par
%%%%%%%%%%%%%%%%%%%%%%%%%%%%%%%%%%%%%%%%%%%%%%%%%%%%%%%%%%%%%%%%%%%%%%%%%%%%%%%
{\bf Theorem 1.}
{\it For any two-qubit state $\sigma$, the positive part ($P$) of
$\sigma^{T_B}$ is a PPT state.
Further, if $\sigma$ is entangled, the partial transposition of the positive
part ($P^{T_B}$) is full rank.}
\par
%%%%%%%%%%%%%%%%%%%%%%%%%%%%%%%%%%%%%%%%%%%%%%%%%%%%%%%%%%%%%%%%%%%%%%%%%%%%%%%
It has been shown that the partial transposition of any
separable state of rank 2 is also rank 2 \cite{Sanpera98a},
and it is obvious that the partial
transposition of any pure separable state is also 
a pure separable state.
Therefore, the fact that $P^{T_B}$ is separable and full rank implies
that $P$, which is rank deficient, must be rank 3, and we obtain 
the following corollary.
\par
%%%%%%%%%%%%%%%%%%%%%%%%%%%%%%%%%%%%%%%%%%%%%%%%%%%%%%%%%%%%%%%%%%%%%%%%%%%%%%%
{\bf Corollary 1.}
{\it For any entangled state $\sigma$ in two qubits, the positive part ($P$)
of $\sigma^{T_B}$ is rank 3.}
\par
%%%%%%%%%%%%%%%%%%%%%%%%%%%%%%%%%%%%%%%%%%%%%%%%%%%%%%%%%%%%%%%%%%%%%%%%%%%%%%%
Theorem 1 states that, in some sense, the partial transposition
in two qubits can also give separable approximations of the entangled states
as well as the best separable approximation \cite{Lewenstein98a},
the closest disentangled state in the relative entropy
measure \cite{Vedral97a}, and so on.
Every entangled state in two qubits can be decomposed into the
separable approximation $\frac{1}{1+\lambda}P^{T_B}$ (normalized)
and the deviation from it [$(|\psi\rangle\langle\psi|)^{T_B}$].
\par
%%%%%%%%%%%%%%%%%%%%%%%%%%%%%%%%%%%%%%%%%%%%%%%%%%%%%%%%%%%%%%%%%%%%%%%%%%%%%%%
Further, it is important to discuss the geometrical meaning of $|\phi\rangle$
of Eq.\ (\ref{eq: Witness}).
In the case of $p_0\!=\!1/2$, $P^{T_B}$ of
Eq.\ (\ref{eq: PT of Normal Form}) becomes positive semidefinite (rank 3)
and it can be seen that $|\phi\rangle$ satisfies
$P^{T_B}|\phi\rangle\!=\!0$.
Geometrically, $P^{T_B}$ of rank 3 is just located at the crossing point of two
edges as shown in Fig.\ \ref{fig: Crossing Point}.
The hyper-plane corresponding to the entanglement witness of
$(|\psi\rangle\langle\psi|)^{T_B}$ is in contact with the PPT ball at the
crossing point, and 
$P^{T_B}\!-\!\lambda(|\psi\rangle\langle\psi|)^{T_B}$
is located in the direction perpendicular to this hyper-plane
(since the entanglement witness plays the role of the normal vector).
In addition, the witness for the nonpositivity $|\phi\rangle\langle\phi|$,
which is the eigenstate of $P^{T_B}$ for a zero-eigenvalue,
also specifies a hyper-plane
which is in contact with the positive ball at the crossing point.
What we showed in the proof of Theorem 1 is that these two hyper-planes
always cross with {\it shallow} angles so that the nonpositivity of
$P^{T_B}\!-\!\lambda(|\psi\rangle\langle\psi|)^{T_B}$
is always detected by the hyper-plane specified by $|\phi\rangle\langle\phi|$.
The inner product of two normal vectors of the hyper-planes
corresponds to
$\hbox{Tr}(|\psi\rangle\langle\psi|)^{T_B}|\phi\rangle\langle\phi|$,
which was shown always to be positive.
\par
%%%%%%%%%%%%%%%%%%%%%%%%%%%%%%%%%%%%%%%%%%%%%%%%%%%%%%%%%%%%%%%%%%%%%%%%%%%%%%%
\begin{figure}
\centerline{\scalebox{0.44}[0.44]{\includegraphics{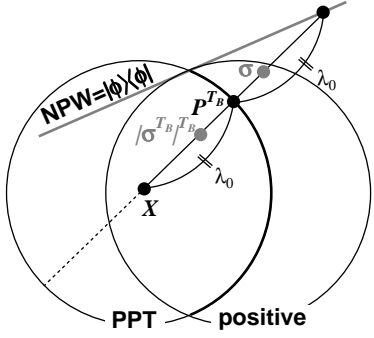}}}
\caption{
An upper bound $\lambda_0$ is obtained so that nonpositivity of 
$\sigma$ is not detected by $|\phi\rangle\langle\phi|$.
$X$ is located apart from $P^{T_B}$ by $\lambda_0$
on opposite side to $\sigma$.
}
\label{fig: Location of X}
\end{figure}
%%%%%%%%%%%%%%%%%%%%%%%%%%%%%%%%%%%%%%%%%%%%%%%%%%%%%%%%%%%%%%%%%%%%%%%%%%%%%%%
The remaining task for the proof of $|\sigma^{T_B}|^{T_B}\!\ge\!0$ is to
search for a positive operator $X$.
According to Lemma 1 and Corollary 1, $P$ and $P^{T_B}$ are represented by
Eqs.\ (\ref{eq: Normal Form}) and 
(\ref{eq: PT of Normal Form}), respectively, when
$\sigma$ is an entangled state.
Further, according to $P^{T_B}\!>\!0$ (Theorem 1),
$p_0\!<\!1/2$ (and $p_3\!=\!0$ since $P$ is rank 3).
So that 
$\sigma\!=\!P^{T_B}-\lambda(|\psi\rangle\langle\psi|)^{T_B}\!\ge\!0$,
the range of $\lambda$ is limited, and an upper bound of
$\lambda$ must be found.
It is slightly surprising that the hyper-plane of
$|\phi\rangle\langle\phi|$ also plays a crucial role for this purpose.
Using $|\phi\rangle$ of Eq.\ (\ref{eq: Witness})
and $\langle \phi|\sigma|\phi\rangle$
of Eq.\ (\ref{eq: Expectation}) (but $p_0\!<\!1/2$ here),
the condition of $\langle \phi|\sigma|\phi\rangle\!\ge\!0$
leads to $\lambda\!\le\!\lambda_0\!\equiv\!(1-2p_0)M/N$
(we again used $\hbox{Tr}CC^*\!\ge\!2$).
Then, we define the operator $X$ as
%%%%%%%%%%%%%%%%%%%%%%%%%%%%%%%%%%%%%%%%%%%%%%%%%%%%%%%%%%%%%%%%%%%%%%%%%%%%%%%
\begin{equation}
X\equiv P^{T_B}+\lambda_0 (|\psi\rangle\langle\psi|)^{T_B},
\end{equation}
%%%%%%%%%%%%%%%%%%%%%%%%%%%%%%%%%%%%%%%%%%%%%%%%%%%%%%%%%%%%%%%%%%%%%%%%%%%%%%%
whose geometrical location is shown in Fig.\ \ref{fig: Location of X}.
This $X$ is always positive as shown below.
Let us introduce
%%%%%%%%%%%%%%%%%%%%%%%%%%%%%%%%%%%%%%%%%%%%%%%%%%%%%%%%%%%%%%%%%%%%%%%%%%%%%%%
\begin{eqnarray}
X'&=& 2N
( \tilde A^\dagger \otimes \tilde B^T) X (\tilde A \otimes \tilde B^*) \cr
&=& 2 
\sum_{i=0}^2
(p_0-p_{3-i})|e_i \rangle\langle e_i| \cr
&+&(1-2p_0) \big[ I \otimes I
+(H_1 \otimes H_2) V (H_1 \otimes H_2)\big] \cr
&=& 2 
\sum_{i=0}^2
(p_0-p_{3-i})|e_i \rangle\langle e_i| \cr
&+&(1-2p_0) (C \otimes I)
\big[ \tilde C^\dagger \tilde C \otimes I
+ V \big] (C^\dagger \otimes I).
\end{eqnarray}
%%%%%%%%%%%%%%%%%%%%%%%%%%%%%%%%%%%%%%%%%%%%%%%%%%%%%%%%%%%%%%%%%%%%%%%%%%%%%%%
Since $A$ and $B$ are full rank, $X\!\ge\!0$ if and only if $X'\!\ge\!0$.
The first term of $X'$ is positive since $p_0$ is largest.
According to Ref.\ \cite{Lewenstein98a},
for a given $R\!\ge\!0$, if $|\xi\rangle$ belongs to the range of $R$
and $\kappa\!\le\!\frac{1}{\langle\xi|R^{-1}|\xi\rangle}$,
then $R\!-\!\kappa|\xi\rangle\langle\xi|\!\ge\!0$.
Since $\det (\tilde C^\dagger \tilde C)\!=\!1$, the eigenvalues of
$\tilde C^\dagger \tilde C$ are written as $\{t,1/t\}$, and we obtain
%%%%%%%%%%%%%%%%%%%%%%%%%%%%%%%%%%%%%%%%%%%%%%%%%%%%%%%%%%%%%%%%%%%%%%%%%%%%%%%
\begin{eqnarray}
\langle\psi^-|[(\tilde C^\dagger \tilde C+I)\otimes I]^{-1}|\psi^-\rangle
&=&\textstyle\frac{1}{2}\hbox{Tr}
(\tilde C^\dagger \tilde C+I)^{-1} \cr
&=&\textstyle\frac{1}{2}(\frac{1}{t+1}+\frac{1}{1/t+1})
=\textstyle\frac{1}{2},
\end{eqnarray}
%%%%%%%%%%%%%%%%%%%%%%%%%%%%%%%%%%%%%%%%%%%%%%%%%%%%%%%%%%%%%%%%%%%%%%%%%%%%%%%
and 
$\tilde C^\dagger \tilde C \!\otimes\! I\!+\! V
\!=\! 
(\tilde C^\dagger \tilde C\!+\!I) \!\otimes\! I
\!-\!2|\psi^-\rangle\langle\psi^-|
\!\ge\!0$.
As a result, since $p_0\!<\! 1/2$, the second term of $X'$ is also positive
and $X'$ is found to be positive.
Since $0\!<\!\lambda\!\le\!\lambda_0$, $|\sigma^{T_B}|^{T_B}$
can be always written as a convex sum of
two positive operators ($X$ and $P^{T_B}$), and
the following theorem was proven.
\par
%%%%%%%%%%%%%%%%%%%%%%%%%%%%%%%%%%%%%%%%%%%%%%%%%%%%%%%%%%%%%%%%%%%%%%%%%%%%%%%
{\bf Theorem 2.}
{\it $|\sigma^{T_B}|^{T_B}\!\ge\!0$ for any two-qubit state $\sigma$.}
\par
%%%%%%%%%%%%%%%%%%%%%%%%%%%%%%%%%%%%%%%%%%%%%%%%%%%%%%%%%%%%%%%%%%%%%%%%%%%%%%%
Finally, we briefly discuss the case in the higher dimensional systems.
It has been already mentioned that
$|\sigma^{T_B}|^{T_B}\!\ge\!0$ does not hold in general,
and
the states violating $|\sigma^{T_B}|^{T_B}\!\ge\!0$ have been named binegative
states \cite{Audenaert02a}.
In order to obtain some insights into how the binegative states emerge,
we numerically generated the random binegative states of full rank in two
qutrits, and confirmed that
$P^{T_B}\!\equiv\!\sigma/2\!+\!|\sigma^{T_B}|^{T_B}/2$
is not positive in general.
This implies that the necessary condition corresponding to Theorem 1 is
already violated in the higher dimensional systems
(binegative states satisfying Theorem 1 also exist).
This seems to imply that the two hyper-planes at the crossing
point (like those shown in Fig.\ \ref{fig: Crossing Point}) sometimes cross
with {\it steep} angles
(it was shown to be always {\it shallow} in two qubits).
\par
%%%%%%%%%%%%%%%%%%%%%%%%%%%%%%%%%%%%%%%%%%%%%%%%%%%%%%%%%%%%%%%%%%%%%%%%%%%%%%%
It will be important to clarify the geometry of the state space more,
which might lead to a geometrical understanding of the quantum
information tasks.
%%%%%%%%%%%%%%%%%%%%%%%%%%%%%%%%%%%%%%%%%%%%%%%%%%%%%%%%%%%%%%%%%%%%%%%%%%%%%%%
%%
%%%%%%%%%%%%%%%%%%%%%%%%%%%%%%%%%%%%%%%%%%%%%%%%%%%%%%%%%%%%%%%%%%%%%%%%%%%%%%%
%\bibliography{personal}
%\bibliographystyle{apsrev}

%%%%%%%%%%%%%%%%%%%%%%%%%%%%%%%%%%%%%%%%%%%%%%%%%%%%%%%%%%%%%%%%%%%%%%%%%%%%%%%
%%
%%%%%%%%%%%%%%%%%%%%%%%%%%%%%%%%%%%%%%%%%%%%%%%%%%%%%%%%%%%%%%%%%%%%%%%%%%%%%%%
\end{document}